\documentclass{article}
\pdfoutput=1
\usepackage[onehalfspacing]{setspace}
\usepackage{fullpage}
\usepackage{sectsty} \sectionfont{\large}
\usepackage{amsmath,amsfonts}
\usepackage{url,verbatim,listings,graphicx,listings}
\usepackage{natbib}

\def\eatzero#1{\ifnum#1=0\relax\else{#1}\fi}
\def\eattwo#1#2{\relax}
\def\svndate$#1: #2-#3-#4 #5 #6 (#7) ${\def\fmtdate{#2--#3--#4}}
\svndate$Date: 2010-03-11 23:29:44 -0500 (Thu, 11 Mar 2010) $

\DeclareMathOperator{\Exponential}{Exponential}
\DeclareMathOperator*{\var}{var}
\DeclareMathOperator{\chud}{chud}
\DeclareMathOperator{\drawfrom}{\text{draw from}}
\newcommand{\R}{\mathbb{R}}

\title{Technical Report No. 1002\\
Department of Statistics, University of Toronto\\
Covariance-Adaptive Slice Sampling}
\author{Madeleine Thompson and Radford M. Neal}
\date{\fmtdate}

\begin{document}

\begin{center}

{\small Technical Report No.\ 1002,
 Department of Statistics, University of Toronto}

\vspace*{0.6in}

{\Large\bf Covariance-Adaptive Slice Sampling}\\[16pt]

\begin{tabular}{cc}
{\large Madeleine Thompson} & {\large Radford M. Neal} \\[3pt]
 Department of Statistics & Departments of Statistics and Computer Science \\
 University of Toronto & University of Toronto \\
 \url{mthompson@utstat.toronto.edu} & \url{radford@utstat.toronto.edu} \\
 & \url{http://www.cs.utoronto.ca/~radford/} \\
  \hspace{3in} & \hspace{3in}
\end{tabular}

March 8, 2010

\end{center}


\begin{abstract}
\noindent We describe two slice sampling methods for taking
multivariate steps using the crumb framework.  These methods use
the gradients at rejected proposals to adapt to the local curvature
of the log-density surface, a technique that can produce much better
proposals when parameters are highly correlated.  We evaluate our
methods on four distributions and compare their performance to that
of a non-adaptive slice sampling method and a Metropolis method.
The adaptive methods perform favorably on low-dimensional target
distributions with highly-correlated parameters.
\end{abstract}

\section{Introduction}

Markov Chain Monte Carlo methods often perform poorly when parameters
are highly correlated.  Our goal has been to develop MCMC methods
that work well on such distributions without requiring prior knowledge
about the distribution or extensive tuning runs.

Slice sampling \citep{neal03} is an auxiliary-variable MCMC method
based on the idea of drawing points uniformly from underneath the
density surface of a target distribution.  If one discards the
density coordinate from the sample, the resulting marginal
distribution is the target distribution.

This document presents two samplers in the ``crumb'' framework
\citep[\S5.2]{neal03}, a general framework for slice sampling methods
that take multivariate steps.  Unlike many MCMC samplers, they
perform well when the target distribution has highly correlated
parameters.  These methods can improve on univariate slice sampling
in the same way Metropolis with a properly chosen multivariate
proposal distribution can improve on Metropolis with a spherical
proposal distribution and on Metropolis updates of one coordinate
at a time.

\section{Multivariate Steps in the Crumb Framework}

\begin{figure}
\begin{center}
\begin{tabular}{cc}
\includegraphics{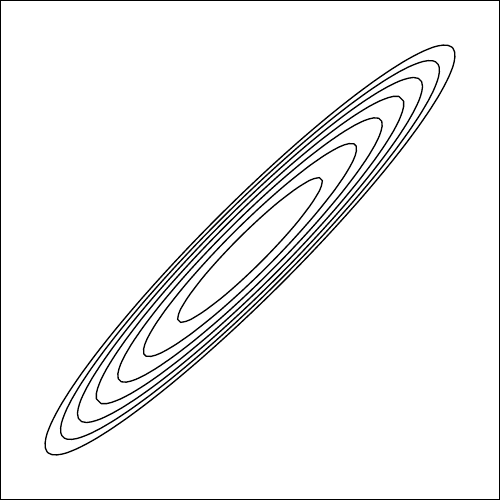} &
\includegraphics{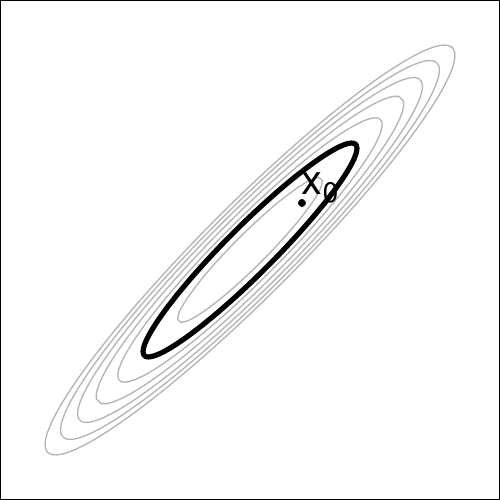} \\
(a) & (b)
\end{tabular}
\end{center}
\caption{(a) Contours of a target distribution.
(b) A slice, $S_{y_0}$, and the current target component, $x_0$.}
\label{ov-contour-slice}
\end{figure}

This section describes the crumb framework for taking multivariate
steps in slice sampling.  The overall goal is to sample from a
target distribution, such as the one with log-density contours shown
in figure~\ref{ov-contour-slice}(a).  The dimension of the target
distribution's parameter space is denoted by $p$.  (The example in
the figure has $p=2$.)  The state space of the Markov chain for
slice sampling has dimension $p+1$, of which $p$ correspond to the
target distribution and one to the current slice level.

Slice sampling alternates between sampling the target coordinates
and the density coordinate.  Let $f(x)$ be proportional to the
density function of
the target distribution, and let $(x_0,y_0)$ be the current state
in the augmented sample space, where $x_0 \in \R^p$ and $y_0 \in
\R$.   To update the density component, $y_0$, we sample uniformly
from $[0,f(x_0)]$.  Updating the target component, $x_0$, is more
difficult.  Let $S_{y_0}$ be the slice through the distribution at
level $y_0$: \begin{equation}S_{y_0} = \{x: f(x) \geq y_0\}\end{equation}
The set $S_{y_0}$ is outlined by a thick line in
figure~\ref{ov-contour-slice}(b).  The difficulty in updating the
target component is that we rarely have an analytic expression for
the boundary of $S_{y_0}$, which may not even be a connected curve,
so we cannot sample uniformly from $S_{y_0}$ as we would like to.
The methods of this document instead perform updates that leave the
uniform distribution on $S_{y_0}$ invariant, leaving the resulting
chain with the desired stationary distribution.

\begin{figure}
\begin{center}
\begin{tabular}{cc}
\includegraphics{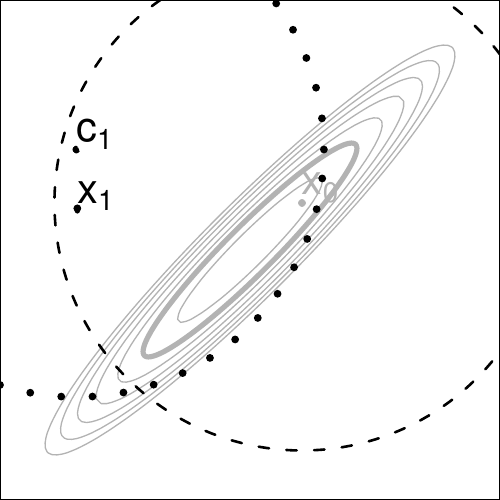} &
\includegraphics{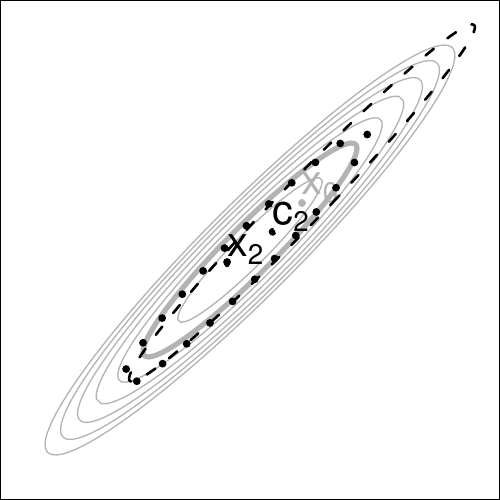} \\
(a) & (b)
\end{tabular}
\end{center}
\caption{(a) The first crumb, $c_1$, and a proposal, $x_1$.  The
distribution of $c_1$ is shown as dashed.  The distribution of $x_1$
is shown as dotted.  In this figure and subsequent ones, probability
distributions in a two-dimensional space are represented by ellipses
such that a uniform distribution over the ellipse has the same mean
and covariance as the represented distribution.  This method allows
multiple distributions to be drawn in the same plot.  (b) The second
crumb, $c_2$, and a proposal, $x_2$.  The distribution of $c_2$ is
shown as dashed.  The distribution of $x_2$ is shown as dotted.
The distribution for $c_2$ has been updated using the method described in
section~\ref{adaptingcovariance}.}
\label{ov-ck-xk}
\end{figure}

These methods begin by sampling a ``crumb,'' $c_1$, from a distribution
that depends on $x_0$, then drawing a proposal, $x_1$, from the distribution
of points that could have generated $c_1$, treating the crumb as
data and $x_0$ as an unknown.  An example crumb and proposal are
drawn in figure~\ref{ov-ck-xk}(a), with the distribution of $c_1$
shown as a dashed line and the distribution of $x_1$ shown as a
dotted line.

If $x_1$ were inside $S_{y_0}$, we would accept it as the new value
for the target component of the state.  In figure~\ref{ov-ck-xk}(a),
it is not, so we draw a new crumb, $c_2$, and draw a new proposal,
$x_2$, from the distribution of $x_0$ given both $c_1$ and $c_2$
as data.  This is plotted in figure~\ref{ov-ck-xk}(b).  While
the distribution of proposed moves is determined by the crumbs and
their distributions, we can choose the distribution of the crumbs
arbitrarily, using the densities and gradients at rejected proposals
if we desire.

In the example of figure~\ref{ov-ck-xk}(b), $x_2$ is in $S_{y_0}$,
so we accept $x_2$ as the new target component.  If it were not,
we would keep drawing crumbs, adapting their distributions so that
the proposal distribution would be as close as possible to uniform
sampling over $S_{y_0}$, and metaphorically following these crumbs
back to $x_0$ by drawing proposals from the distribution of $x_0$
conditional on having observed the crumbs \citep{grimm}.

\citet[\S5.2]{neal03} has more information on the crumb framework,
including a proof that methods in this framework leave the target
distribution invariant.

\section{Overview of Adaptive Gaussian Crumbs}
\label{adaptivecrumb}

We now specialize the crumb framework to Gaussian crumbs and proposals,
using log densities at proposals and their gradients to choose the
crumb covariances.  Without violating detailed balance, the crumb
distribution can depend on these log densities and gradients.  We
assume that while computing the log density at a proposal, we can
compute its gradient with minimal additional cost.

Ideally, the proposal distribution would be a uniform distribution
over $S_{y_0}$.  To approximate uniform sampling over $S_{y_0}$,
we attempt to draw a sequence of crumbs that results in a Gaussian
proposal distribution with the same covariance as a uniform
distribution over the slice.

In both adaptive methods discussed in this document, the first
crumb has a multivariate Gaussian distribution:
\begin{equation}
c_1 \sim N(x_0,W_1^{-1})
  \quad\text{where $W_1 = \sigma_c^{-2} I$}
\end{equation}
The standard deviation of the initial crumb, $\sigma_c$, is the
only tuning parameter for either method that is modified in normal
use.  The distribution for $x_0$ given $c_1$ is a Gaussian with
mean $c_1$ and precision matrix $W_1$, so we draw a proposal from
this distribution:
\begin{equation}
x_1 \sim N(c_1,W_1^{-1})
\end{equation}
If $f(x_1)$ is at least $y_0$, then $x_1$ is inside the slice, so
we accept $x_1$ as the target component of the next state of the
chain.  This update leaves the density component, $y_0$, unchanged, though
it is usually forgotten after a proposal is accepted, anyway.

When the proposal is not in the slice, we choose a different
covariance matrix, $W_{k+1}$, for the next crumb, so that the
covariance of the next proposal will look more like that of uniform
sampling over $S_{y_0}$.  The two methods proposed in this document
differ in how they make that choice; sections \ref{adaptingcovariance}
and \ref{shrinkingrank} describe the details.

After sampling $k$ crumbs from Gaussians with mean $x_0$ and precision
matrices $(W_1,\ldots, W_k)$, the distribution for the $k$th proposal
(the posterior for $x_0$ given the crumbs) is:
\begin{align}
x_k &\sim N(\bar c_k, \Lambda_k^{-1}) \label{xk} \\
\label{lambdaxk} \text{where}\;
  \Lambda_k &= W_1 + \cdots + W_k \\
\text{and}\; \bar c_k &= \Lambda_k^{-1}
  ( W_1 c_1 + \cdots + W_k c_k ) \label{barck}
\end{align}
If $x_k \in S_{y_0}$---that is, if $f(x_k) \geq y_0$---we accept
$x_k$ as the target component.  Otherwise, we must choose a
covariance for the distribution of $(k+1)$th crumb, draw a new
proposal, and repeat until a proposal is accepted.

\section{First Method: Matching the Slice Covariance}
\label{adaptingcovariance}

In the method described in this section, we attempt to find $W_{k+1}$
so that the $(k+1)$th proposal distribution has the same conditional
variance as uniform sampling from $S_{y_0}$ in the direction of the
gradient of $\log f(\cdot)$ at $x_k$.  This gradient is a good guess
at the direction in which the proposal distribution is least like $S_{y_0}$.
Figure~\ref{ac-cuts}(a) shows an example of this.  We plotted the
gradients of the log density at thirty points drawn from the same
distribution (shown as a dotted line) as a rejected proposal; most
of these gradients point in the direction where the proposal variance is
least like the slice.  Generally, in an ill-conditioned distribution,
these gradients do not point towards the mode, they point towards
the nearest point on the slice.  (In a well-conditioned distribution,
the directions to the mode and to the nearest point on the slice
will be similar.)

\subsection{Choosing Crumb Covariances}
\label{choosingcovar}

\begin{figure}
\begin{center}
\begin{tabular}{ccc}
\includegraphics{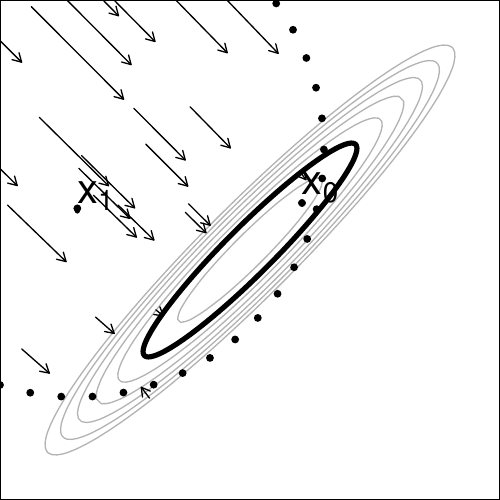} & 
\includegraphics{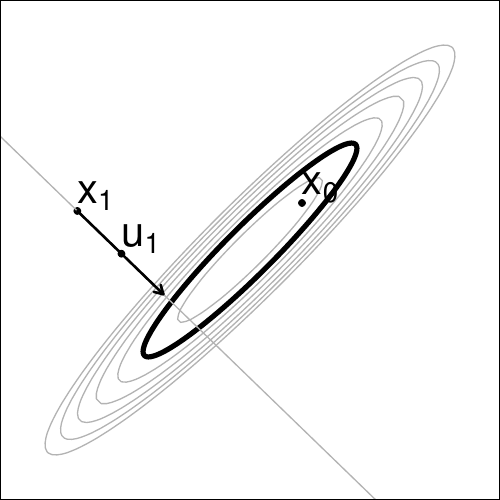} & 
\includegraphics{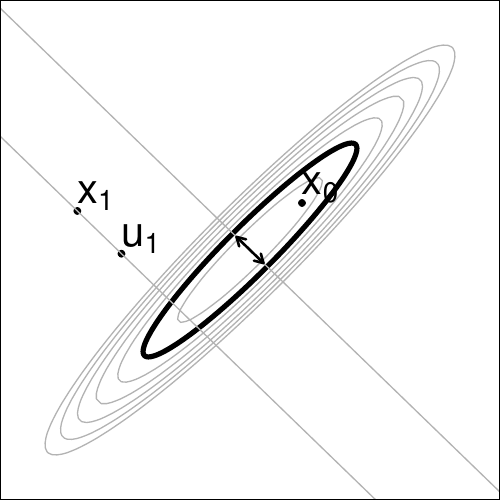} \\
(a) & (b) & (c)
\end{tabular}
\end{center}
\caption{(a) shows gradients at thirty points drawn from the first
proposal distribution (shown as a dotted line).  These gradients
tend to point towards the slice (shown as a thick line).  (b) shows
a parabolic cut through the log-density surface that goes through
a rejected proposal, $x_1$, in the direction of the gradient at
$x_1$, shown as an arrow.  $u_1$ is a point on the cut defined by
equation~\ref{uk}.  (c) adds a parabolic cut through the mode in
the same direction.  These two parabolas have approximately the
same second derivative.  The distance between the two arms of the second
parabola when the vertical coordinate is equal to $\log y_0$ is shown
as a double-ended arrow, equal to $d$ in equation~\ref{d}.}
\label{ac-cuts}
\end{figure}

To compute $W_{k+1}$, we will estimate the second derivative of the target
distribution in the direction of the gradient, assuming the target
distribution is approximately locally Gaussian.  Consider the
(approximately) parabolic cut through the log-density surface in
figure~\ref{ac-cuts}(b), which has the equation:
\begin{equation}\label{paraboliccut}
\ell = -\frac12 \kappa t^2 + \beta t + \gamma
\end{equation}
$t$ is a parameter that is zero at $x_k$ and increases in the
direction of the gradient; $\kappa$, $\beta$, and $\gamma$ are
unknown constants we wish to estimate.  The coefficient $-\frac12$
is arbitrary; this choice makes $\kappa$ equal to the negative of
the second derivative of the parabola.  We already had to compute
$f(x_k)$ so that we could determine whether $x_k$ was in $S_{y_0}$;
we assume that $\nabla \log f(x_k)$ was computed simultaneously.
To fix two degrees of freedom of the parabola, we plug these
quantities into equation~\ref{paraboliccut} and its derivative with
respect to $t$:
\begin{align}
\label{kappa1}
  \log f(x_k) &= -\frac12 \kappa \cdot 0^2 + \beta \cdot 0 + \gamma 
  = \gamma \\
\label{kappa2} \lVert \nabla \log f(x_k) \rVert &= -\kappa \cdot 0 + \beta 
  = \beta
\end{align}
We still have one degree of freedom left, so we must evaluate $\log
f(\cdot)$ at another point on the parabola.  We choose a point
as far away from $x_k$ as $c_k$ is, hoping that this distance is
within the range where the distribution is approximately Gaussian.
Let $\delta$ be this distance:
\begin{equation}
\delta = \lVert x_k - c_k\rVert
\end{equation}
Let $g$ be the normalized gradient at $x_k$:
\begin{equation}
g=\frac{\nabla \log f(x_k)}{\lVert \nabla \log f(x_k) \rVert}
\end{equation}
Then, the point $u_k$ is defined to be $\delta$ away from $x_k$ in
the direction $g$:
\begin{equation}\label{uk}
u_k = x_k + \delta \cdot g
\end{equation}
In equation~\ref{paraboliccut}, $u_k$ corresponds to $t=\delta$.
We evaluate the density at $u_k$ to fix the parabolic cut's
third degree of freedom, and plug this into equation~\ref{paraboliccut}:
\begin{equation}\label{kappa3}
  \log f(u_k) = -\frac12 \kappa \delta^2 + \beta \delta + \gamma
\end{equation}
Solving equations \ref{kappa1}, \ref{kappa2}, and \ref{kappa3}
for $\kappa$ gives:
\begin{equation}\label{kappax1}
  \kappa = -\frac{2}{\delta^2} \bigl( \log f(u_k) - \log f(x_k)
    - \lVert \nabla \log f(x_k) \rVert \cdot \delta \bigr)
\end{equation}

We can now use $\kappa$ to approximate the conditional variance
in the direction $g$.  If the Hessian is locally approximately
constant, as it is for Gaussians, a cut through the mode in the
direction of $\nabla \log f(x_k)$ would have the same second derivative
as the cut through $x_k$.  This second parabola, shown in
figure~\ref{ac-cuts}(c), has the equation:
\begin{equation}\label{modeparabola}
\ell = -\frac12 \kappa t^2 + M
\end{equation}
$M$ is the log density at the mode.  We set $t=0$ at the mode, so
there is no linear term.  For now, assume $f(\cdot)$ is unimodal
and that $M$ was computed with the conjugate gradient method
\citep[ch.~5]{nocedal06} or some other similar procedure before
starting the Markov chain.  We solve equation~\ref{modeparabola}
for the parabola's diameter $d$ at the level of the current slice,
$\log y_0$, shown as a doubled-ended arrow in figure~\ref{ac-cuts}(c):
\begin{equation}\label{d}
d = \sqrt{\frac{8(M-\log y_0)}{\kappa}}
\end{equation}
Since the distribution of points drawn from an ellipsoidal slice,
conditional on their lying on that particular one-dimensional cut, is
uniform with length $d$, the conditional variance in the direction
of the gradient at $x_k$ is:
\begin{equation}
\sigma^2_{k+1} = \frac{d^2}{12} = \frac23 \cdot \frac{M-\log y_0}{\kappa}
\end{equation}

With this variance, we can construct a crumb precision matrix
that will lead to the desired proposal precision matrix.  We want to
draw a crumb $c_{k+1}$ so that the posterior of $x_0$ given the $k$
crumbs has a variance equal to $\sigma_{k+1}^2$ in the direction
$g$.  Using equation~\ref{lambdaxk}, the
precision of the proposal given these crumbs is:
\begin{equation}\label{lambdak+1}
\Lambda_{k+1} = \Lambda_k + W_{k+1}
\end{equation}
If we multiply both sides of equation~\ref{lambdak+1} by $g^T$ on
the left and $g$ on the right, the left side is the conditional
precision in the direction $g$.
\begin{equation}\label{projxk+1}
g^T\Lambda_{k+1}g = g^T\bigl(\Lambda_k + W_{k+1} \bigr) g
\end{equation}
We would like to choose $W_{k+1}$ so that this conditional precision
is $\sigma_{k+1}^{-2}$, so we replace the left hand side of
equation~\ref{projxk+1} with that:
\begin{equation}\label{projxk+1b}
\sigma_{k+1}^{-2} = g^T\bigl(\Lambda_k + W_{k+1}\bigr)g
\end{equation}
As will be discussed in section~\ref{crumbcovar}, computation will
be particularly easy if we choose $W_{k+1}$ to be a scaled copy of
$\Lambda_k$ with a rank-one update, so we choose $W_{k+1}$ to be
of the form:
\begin{equation}\label{Wk+1}
W_{k+1} = \theta \Lambda_k + \alpha gg^T
\end{equation}
$\alpha$ and $\theta$ are unknown scalars.  $\theta$ controls how
fast the precision as a whole increases.  If we would like the
variance in directions other than $g$ to shrink by $9/10$, for
example, we choose $\theta = 1/9$.  Since $\theta$ is constant,
there will be exponentially fast convergence of the proposal
covariance in all directions, which allows quick recovery from
overly diffuse initial crumb distributions.  For this method,
$\theta$ is not a critical choice; $\theta=1$ is reasonable.
Substituting equation~\ref{Wk+1} into equation~\ref{projxk+1b}
gives:
\begin{equation}
\sigma_{k+1}^{-2} =
  g^T\bigl(\Lambda_k + \theta\Lambda_k+\alpha gg^T\bigr)g
\end{equation}
Noting that $g^Tg = 1$, we solve for $\alpha$, and set:
\begin{equation}\label{alpha}
\alpha = \max\{ \sigma_{k+1}^{-2}-(1+\theta)g^T\Lambda_kg , 0\}
\end{equation}
$\alpha$ is restricted to be positive to guarantee positive
definiteness of the crumb covariance.  (By choosing $\theta$
simultaneously, we could perhaps encounter this restriction less
frequently, but we have not explored this.)  Once we know $\alpha$,
we then compute $W_{k+1}$ using equation~\ref{Wk+1}.

The resulting crumb distribution is:
\begin{equation}\label{ck+1}
c_{k+1} \sim N(x_0, W_{k+1}^{-1})
\end{equation}
After drawing such a crumb, we draw a proposal according to
equations~\ref{xk} to \ref{barck}, accepting or rejecting depending
on whether $f(x_{k+1}) \geq y_0$, drawing more crumbs and adapting
until a proposal is accepted.

\subsection{Estimating the Density at the Mode}
\label{estimatingM}

We now modify the method to remove the restriction that the target
distribution be unimodal and remove the requirement that we precompute
the log density at the mode.  Estimating $M$ each time we update
$x_0$ instead of precomputing it allows $M$ to take on values
appropriate to local modes.  Since the proposal distribution only
approximates the slice even in the best of circumstances, it is not
essential that the estimate of $M$ be particularly good.

To estimate $M$, we initialize $M$ to the slice level, $\log y_0$, before
drawing the first crumb.  Then, every time we fit the parabola
described by equation~\ref{paraboliccut}, we update $M$ to be the
maximum of the current value of $M$ and the estimated peak of the
parabola.  As more crumbs are drawn, $M$ becomes a better estimate
of the local maximum.  We could also use the values of the log
density at rejected proposals and at the $\{u_k\}$ to bound $M$,
but if the log density is locally concave, the log densities at the
peaks of the parabola will always be larger than these values.

\subsection{Efficient Computation of the Crumb Covariance}
\label{crumbcovar}

This section describes a method for using Cholesky factors of
precision matrices to make implementation of the covariance-matching
method of section~\ref{choosingcovar} efficient.  If implemented
naively, the method of section~\ref{choosingcovar} would use $O(p^3)$
operations when drawing a proposal with equations~\ref{xk} and
\ref{barck}.  We would like to avoid this.  One way is to represent
$W_k$ and $\Lambda_k$ by their upper-triangular Cholesky factors
$F_k$ and $R_k$, where $F_k^T F_k = W_k$ and $R_k^T R_k = \Lambda_k$.

First, we must draw proposals efficiently.  If $z_1$ and $z_2$ are
$p$-vectors of standard normal variates, we can replace the crumb
and proposal draws of equations \ref{ck+1} and \ref{xk} with:
\begin{align}
\label{ck2} c_k &= x_0 + F_k^{-1} z_1 \\
\label{xk2} x_k &= \bar c_k + R_k^{-1} z_2
\end{align}
Since Cholesky factors are upper-triangular, evaluation of $F_k^{-1}
z_1$ and $R_k^{-1} z_2$ by backward substitution takes $O(p^2)$ operations.

We must also update the Cholesky factors efficiently.  We replace
the updates of $W_k$ and $\Lambda_k$ in equations~\ref{Wk+1}
and~\ref{lambdak+1} with:
\begin{align}
F_{k+1} &= \chud(\sqrt{\theta}R_k, \sqrt{\alpha} g) \\
R_{k+1} &= \chud(\sqrt{1+\theta} R_k, \sqrt{\alpha} g)
\end{align}
Here, $\chud(R, v)$ is the Cholesky factor of $R^TR+vv^T$.  The
function name is an abbreviation for ``Cholesky update.''  It can
be computed with the LINPACK routine \texttt{DCHUD}, which uses
Givens rotations to compute the update in $O(p^2)$ operations
\citep[ch.~10]{linpack}.

Finally, we would like to compute the proposal mean efficiently.
We do this by keeping a running sum of the un-normalized crumb mean
(the parenthesized expression in equation~\ref{barck}), which we will
represent by $\bar c_k^*$.  Define:
\begin{align}
\bar c_k^* &= W_1 c_1 + \cdots + W_k c_k \\
&= \bar c_{k-1}^{*} + W_k c_k \\
&= \bar c_{k-1}^* + F_k^T F_k c_k
\end{align}
Then, using forward and backward substitution, we can compute the
normalized crumb mean, $\bar c_k$, as:
\begin{equation}
\bar c_k = R_k^{-1} R_k^{-T} \bar c_k^*
\end{equation}
This way, we can compute $\bar c_k$ in $O(p^2)$ operations and do
not need to save all the crumbs and crumb covariances.

With these changes, the resulting algorithm is numerically stable
even with ill-conditioned target distributions.  Each crumb and
proposal draw takes $O(p^2)$ operations.  Figure~\ref{adaptivecrumbpseudo}
shows pseudocode for this method.

\begin{figure}
\input{crumb-fig.tex}
\label{adaptivecrumbpseudo}
\end{figure}

\section{Second Method: Shrinking the Rank of the Crumb Covariance}
\label{shrinkingrank}

The method of section~\ref{adaptingcovariance} attempts to adapt the
crumb distribution so that the proposal distribution matches the
shape of the slice.  However, it often can't due to positive-definiteness
constraints, requiring the $\max\{\cdot,\cdot\}$ operation in
equation~\ref{alpha}.  Even when it can perform the adaptation, it
may not be appropriate if the underlying distribution is not
approximately Gaussian.

This section describes a different method, also based on the framework
of section~\ref{adaptivecrumb}.  Instead of attempting to match the
conditional variance in the direction of the gradient, it just
sets it to zero.  This is reasonable in that the gradient at a proposal
probably points in a direction where the variance is small, so it
is more efficient to move in a different direction.

With this method, the crumb covariance is zero in some directions
and spherically symmetric in the rest, so its simplest representation
is the pair $(w, J)$, where $J$ is a matrix of orthonormal columns
of directions in which the conditional variance is zero, and $w^2$
is the variance in the other directions.

Define $P(J,v)$ to be the function that returns the component of vector
$v$ orthogonal to the columns of $J$:
\begin{equation}\label{projfunction}
P(J,v) = \begin{cases} v - JJ^T v & \quad \text{if $J$ has at
least one column} \\ v & \quad \text{if $J$ has no columns} \end{cases}
\end{equation}
For simplicity of computation, since the crumbs are located in a
common subspace with $x_0$, this method will consider the origin
to be at $x_0$ except when calling $f(\cdot)$, which is provided
by the user, and when returning samples to the user.  Each crumb
is drawn by:
\begin{equation}
c_k = \sigma_c \, P(J, z) \qquad \text{where $z\sim N_p(0,I)$}
\end{equation}
When the first crumb is drawn, $J$ has no columns, so $P(J,z) = z$
and the first crumb has the distribution specified in
section~\ref{adaptivecrumb}.

Given $k$ crumbs and the covariances of their distributions,
we know that $x_0$ must lie in the intersection of the subspaces of
their covariances.  Since the subspace $c_j$ is drawn
from contains the subspace $c_k$ is drawn from when $k>j$, this is equivalent
to saying that $x_0$ must lie in the subspace of $c_k$'s covariance,
the orthogonal complement of $J$.  So, the precision of the posterior
for $x_0$ in the direction of columns of $J$ is infinite. It is
equal to $k\sigma_c^{-2}$ in all other directions, since there are
$k$ crumbs, each with spherical variance equal to $\sigma_c^2$ in
the subspace they were drawn in.  The mean of the proposal distribution
(with origin $x_0$) is the projection of:
\begin{align}
\bar c
  &= \frac{\sigma_c^{-2} c_1 + \cdots +
    \sigma_c^{-2} c_k}{k\sigma_c^{-2}}\nonumber \\
  &= k^{-1} \left( c_1 + \cdots + c_k \right)
\end{align}
Therefore, to draw a proposal, we draw a vector of standard normal
variates, project it into the orthogonal complement of $J$, scale
by $\sigma_c/\sqrt{k}$, and add $\bar c$, also projected into the
orthogonal complement of $J$.  With the original origin, the proposal
is:
\begin{equation}
x_k = x_0 + P(J, \bar c) + (\sigma_c/\sqrt{k}) \cdot P(J,z)
  \qquad \text{where $z\sim N_p(0,I)$}
\end{equation}
If the proposal is in the slice (that is, if $f(x_k) \geq y_0$),
we accept it.  Otherwise, we adapt the crumb distribution.  If $J$
has $p-1$ columns, we can't add any more without the crumb covariance
having a rank of zero, so we do not adapt in that case.  Otherwise,
we add a single new column in the direction of $\nabla \log f(x_k)$,
projected into the directions not already spanned by $J$, which we
denote by $g^*$.  Thus, the new value of $J$ would be:
\begin{align}
J_{k+1} &= \left[ J_k \quad \frac{g^*}{\lVert g^* \rVert} \right] \\
\text{where}\quad g^* &= P\left(J_k, \nabla \log f(x_k) \right)
\end{align}
To prevent meaningless adaptations, we only perform this operation
when the angle between the gradient and its projection into the
nullspace of $J$ is less than $60^\circ$.
Equivalently, we only adapt when:
\begin{equation}
\frac{{g^*}^T \nabla\log f(x_k)}{\lVert g^* \rVert \,
  \lVert \nabla\log f(x_k) \rVert} > \frac12
\end{equation}
After possibly updating $J$, we draw another crumb and proposal, repeating
until a proposal is accepted.  The method of this section is
summarized in figure~\ref{orthogonalcrumbpseudo}.

\begin{figure}
\input{orth-fig.tex}
\label{orthogonalcrumbpseudo}
\end{figure}

A variation on the method (which we use in the implementation tested
in section~\ref{results}) shrinks the crumb standard deviation in
the nonzero directions, $\sigma_c$, by a constant factor each
time a crumb is drawn; section \ref{results} uses 0.9.  This results
in exponentially falling proposal variance, which, as described in
section~\ref{choosingcovar}, allows large initial crumb variances
to be used.

\section{Procedure for Evaluation}
\label{procedure}

Figures~\ref{results-cor4} and \ref{results-all} demonstrate the
performance of the methods described in this document.
Figure~\ref{results-cor4} demonstrates the performance of four
samplers on a Gaussian distribution (to be described in
section~\ref{results}).  It contains two graphs, one for each of
two figures of merit.  The top graph plots log density function
evaluations per independent sample against a tuning parameter.  The
bottom graph plots processor-seconds per independent sample against
a tuning parameter.  For both figures of merit, smaller is better.

Each of the four panes in each graph contains data from a single
sampler, with each point representing a run with a specific tuning
parameter.  The tuning parameter for each sampler has the same
scale; the sampler initially attempts to take steps roughly that size.

Both figures of merit require us to determine the correlation length,
the number of correlated samples equivalent to an independent sample.
For the runs done here, an AR(1) model captures the necessary
structure.  For each parameter, we fit the following model:
\begin{equation}
X_t = E(X_t) + \pi \cdot X_{t-1} + a_t
\end{equation}
where $a_t$ is a noise process.  Then, the number of samples
equivalent to a single independent sample is:
\begin{equation}
\tau = \frac{\var(a_t)}{\var(X_t) \cdot (1-\pi)^2}
\end{equation}
This formula is based on the \texttt{effectiveSize} function in
CODA \citep{plummer06}, which uses the spectral approach of
\citet{heidelberger81}.  \citet[pp.~276--278]{wei06} has a more
in-depth discussion of the spectrum of AR processes.  To estimate
a correlation length for a multivariate distribution, we take the
largest estimated correlation length of each of its parameters.
This is not valid in general, but is an acceptable approximation
for this experiment.

Most chains are displayed with a circle indicating an estimate of
the figure of merit, with a line indicating a nominal 95\% confidence
interval.  The intervals are based on normality of the increments
of the AR(1) process, so they should be viewed as a lower bound on
the uncertainty of the point estimates.  Chains whose figures are
plotted as question marks were estimated as having an effective
sample size of less than four.

Figure~\ref{results-all} contains information on four different
samplers.  The panes of figure~\ref{results-all} are similar to
those of figure~\ref{results-cor4}, and each is labeled with the
distribution and the sampler the chains in that pane come from.
The columns of panes correspond to samplers; the rows of panes
correspond to distributions.  By reading across, one can see how
different samplers perform on a given distribution.  By reading
down, one can see how the performance of a given sampler varies
from distribution to distribution.

Every chain has 150,000 samples.  In general, the chain length does
not affect the results; we will point out exceptions to this.

\section{Results of Evaluation}
\label{results}

We simulated four samplers on four distributions for twelve different
tuning parameters each.  The samplers we considered are:
\begin{itemize}
\item Covariance-Matching: This is the method described in
section~\ref{adaptingcovariance}.  The tuning parameter is $\sigma_c$.
\item Shrinking-Rank: This is the method described in
section~\ref{shrinkingrank}.  The tuning parameter is $\sigma_c$.
\item Non-Adaptive Crumbs: This is a non-adaptive variant of the
general method of section~\ref{adaptivecrumb}.  It is like
the method of section~\ref{shrinkingrank}, but never shrinks
rank.  However, like that method, it scales the crumb standard
deviation down by $0.9$ after each proposal.  The tuning parameter
is $\sigma_c$.
\item Metropolis (with Trials): This method is a Metropolis sampler
that uses trial runs to automatically determine a suitable Gaussian
proposal distribution.  The tuning parameter specifies the standard
deviation of the proposal distribution for the first trial run.
\end{itemize}
The distributions we considered are:
\begin{itemize}
\item $N_4(\rho=0.999)$:  This is a four-dimensional Gaussian with
highly-correlated parameters.  The marginal variances for each
parameter are one; each parameter has a correlation of $0.999$ with
each other parameter.  The covariance matrix has a condition number of
about 2900 and is not diagonal.
\item $N_4(\rho=-0.3329)$: This is a four-dimensional Gaussian with
negatively-correlated parameters.  The marginal variances for each
parameter are one, and each parameter has a correlation of $-0.3329$
with each other parameter.  The covariance has a condition number
of about 2900, like $N_4(\rho=0.999)$, but instead of one large
eigenvalue and three small ones, this distribution has one small
eigenvalue and three large ones.
\item Eight Schools: This is a multilevel model in ten dimensions,
consisting of eight group means and hyperparameters for their mean
and variance.  It comes from \citet[pp.~138--145]{gelman04}.
\item Ten-Component Mixture: This is a ten-component
Gaussian mixture in $\R^{10}$.  Each mode is a spherically symmetric
Gaussian with unit variance.  The modes are uniformly distributed
on a hypercube with edge-length ten.
\end{itemize}

\begin{figure}[tb]
\begin{center}
\includegraphics{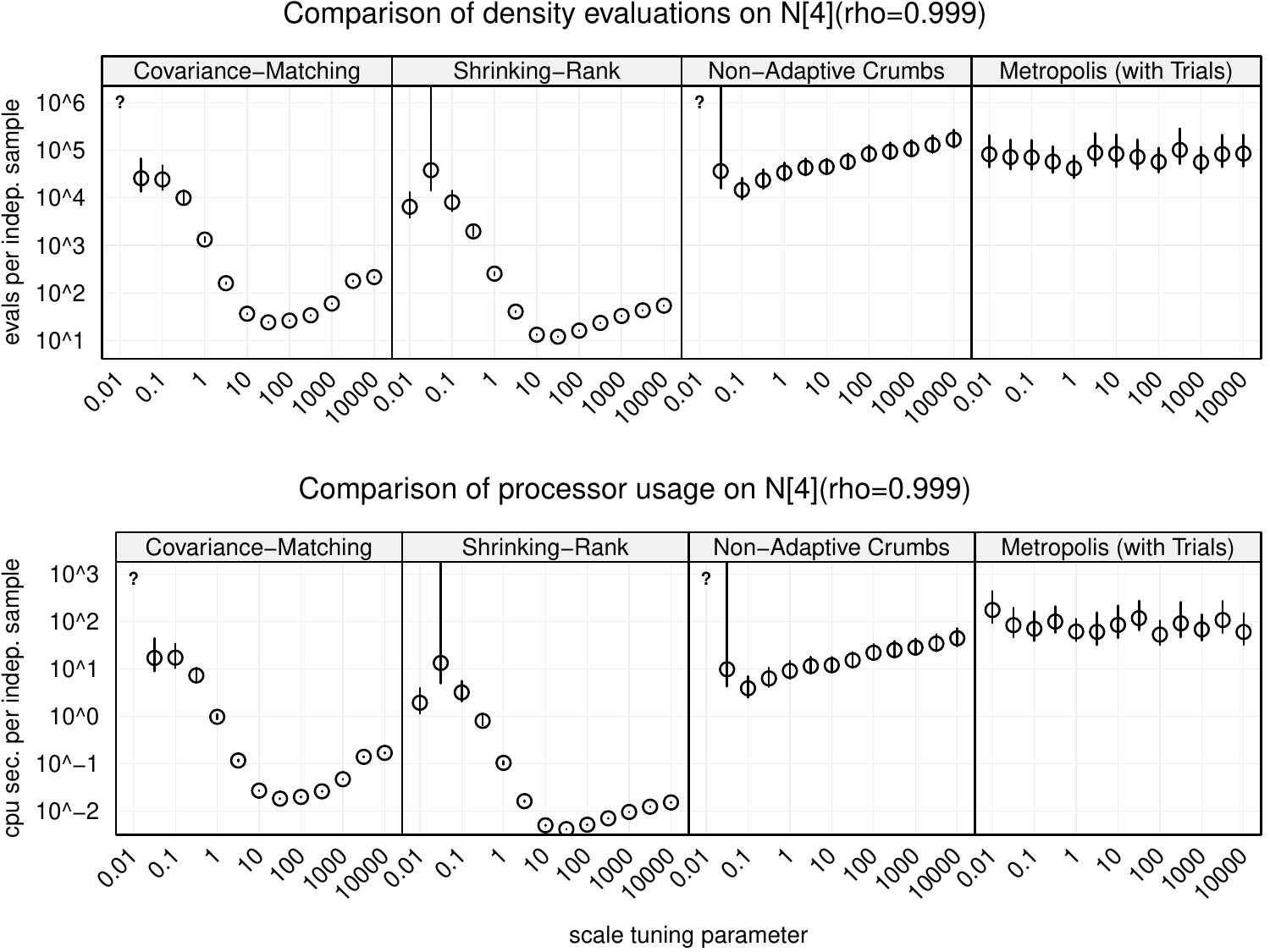}
\end{center}
\caption{The performance of four MCMC samplers on $N_4(\rho=0.999)$.
See section~\ref{procedure} for a description of the graphs and
section~\ref{results} for discussion.}
\label{results-cor4}
\end{figure}

\begin{figure}
\begin{center}
\includegraphics{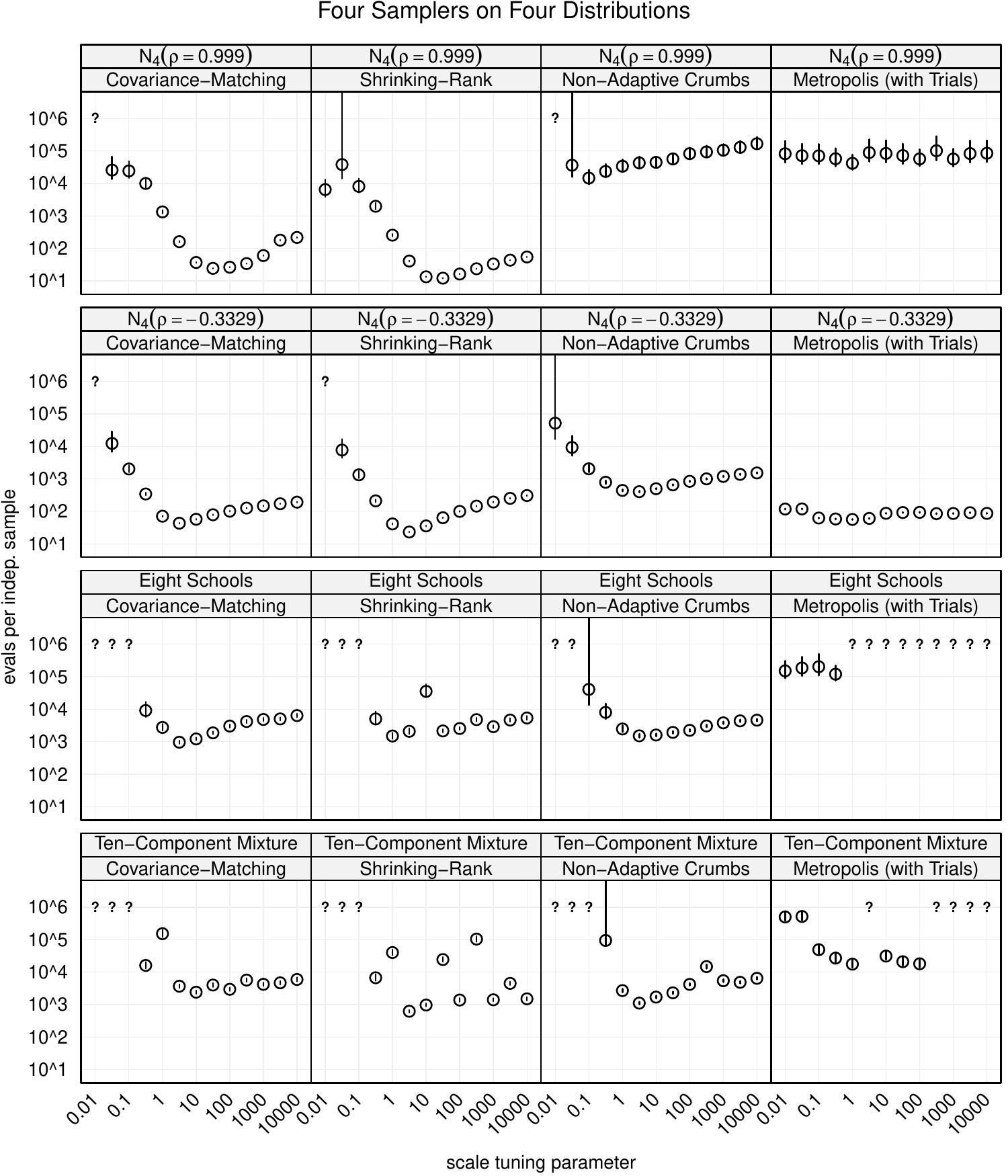}
\end{center}
\caption{The performance of four MCMC samplers on four distributions.
See section~\ref{procedure} for a description of the graph and
section~\ref{results} for discussion.}
\label{results-all}
\end{figure}

By comparing the top and bottom graphs of figure~\ref{results-cor4},
we see that for $N_4(\rho=0.999)$, processor time and number of
density evaluations tell the same story.  The plots of function
evaluations and the plots of processor time are nearly identical
except for their vertical scales.  This is true of the other
distributions as well, so we do not repeat the processor-time plots
for the others.

Figure~\ref{results-cor4} (and the identical first row of
figure~\ref{results-all}) shows the performance of the four methods
on a highly-correlated four-dimensional Gaussian, $N_4(\rho=0.999)$.
Both adaptive slice sampling methods perform well once the tuning
parameter is at least the same order as the standard deviation of
the target distribution.  This hockey-stick performance curve is
characteristic of slice sampling methods.  The non-adaptive slice
sampling method always takes steps of the order of the smallest
eigenvalue once its tuning parameter is at least that large, so its
performance is bad, but after this threshold, its performance does
not depend much on the tuning parameter.  The Metropolis sampler
also fails to capture the shape of the distribution, a result that
depends more on chain length.  Had the chain been longer, the
preliminary runs the sampler uses to estimate a proposal distribution
might have worked better, leading to average performance comparable
to the adaptive slice samplers.

The second row of figure~\ref{results-all} shows sampler performance
on a similar but negatively correlated four-dimensional Gaussian,
$N_4(\rho=-0.3329)$.  The adaptive slice sampling methods continue
to perform well on this distribution, and the non-adaptive sampler
improves somewhat.  The Metropolis sampler improves a great deal;
on this target distribution, it is able to choose a reasonable
proposal distribution, so it performs comparably to the adaptive
slice sampling methods.

The third row of figure~\ref{results-all} shows sampler performance
on Eight Schools.  The minimum threshold appears again for both
adaptive slice samplers as well as the non-adaptive slice sampler.
Since the condition number of the covariance of this distribution
is only about seven, adaptivity is not as important, though slice
sampling's robustness to improper tuning parameters remains important.
The adaptive Metropolis method again fails to identify a reasonable
proposal distribution for small tuning parameters, and fails to
generate any proposal distribution at all for large ones.   This
is partially a reflection on this particular implementation, which
only tries preliminary runs with proposal distribution standard
deviations within four orders of magnitude of the tuning parameter.

The bottom row of figure~\ref{results-all}, which shows performance
on Ten-Component Mixture, has a similar pattern.  The results are
more erratic since the distribution has multiple, moderately-separated
modes, and none of the samplers are designed to perform well on
multimodal distributions.

\section{Discussion}

The adaptive slice sampling methods of sections~\ref{adaptingcovariance}
and~\ref{shrinkingrank} generally perform at least as well as
non-adaptive slice sampling methods and Metropolis.  Slice sampling
in general tends to be more robust to imperfect choice of tuning
parameters than Metropolis.  Preliminary chains are usually
unnecessary, avoiding the hassle of manual management of these runs
or the idiosyncratic performance of automatic evaluation of the
runs.  The main disadvantage of the adaptive slice samplers relative
to Metropolis and non-adaptive slice sampling is that they require
the log density to have an analytically computable gradient, though
this is a standard requirement in numerical optimization, and
experience in that domain has shown that computing the gradient is
often straightforward.

The two adaptive slice sampling methods tend to perform similarly
to each other.  The shrinking-rank method usually performs slightly
better, but this advantage can be mitigated by making the approximation
$\log f(u_k) \approx \log y_0$.  There is no theoretical justification
for this, but it cuts the number of log density evaluations by half
with negligible performance cost, making the performance of the two
adaptive methods indistinguishable.  The shrinking-rank method is
simpler, though, and requires only $O(\min(k,p)\cdot p)$ operations
to draw the $k$th crumb and proposal, slightly better than the
$O(p^2)$ operations needed by the covariance-matching method.

Like most variations of multivariate slice sampling, both the
non-adaptive and adaptive methods described here do not work well
for target distributions in spaces higher than a few dozen.  The
variation in log density of samples increases with dimension, but
slice sampling takes steps in the log density of only order one
each iteration.  So, in high-dimensional spaces, samples tend to be
highly correlated.

Due to this poor scaling with dimensionality, the methods of this
document have limited usefulness on their own.  They may be useful
in highly-correlated low-dimensional problems, though, and can be
used to take steps in highly-correlated low-dimensional subspaces
as part of larger sampling schemes.  We hope to address this
limitation in future work, possibly with polar slice sampling
\citep{roberts02}.

An R implementation of these methods is available at
\url{http://www.cs.utoronto.ca/~radford}.

\bibliographystyle{apalike}
\bibliography{../corlen}

\end{document}